# Tuning of Interlayer Coupling in Large-Area Graphene/WSe$_2$ van der Waals Heterostructure via Ion Irradiation: Optical Evidences and Photonic Applications

Yang Tan[†], Xiaobiao Liu[†], Zhiliang He[‡], Yanran Liu[†], Mingwen Zhao[†]*, Han Zhang[‡]*, and Feng Chen[†]*,

[†]School of Physics, State Key Laboratory of Crystal Materials, Shandong University, Shandong, Jinan, 250100, China,

[‡]Shenzhen Key Laboratory of two dimensional materials and devices, Shenzhen Engineering Laboratory of Phosphorene and Optoelectronics, Key Laboratory of Optoelectronic Devices and Systems of Ministry of Education and Guangdong Province, Shenzhen University, Shenzhen 518060, China

KEYWORDS: Heterostructure, graphene/WSe$_2$, interlayer coupling, ion irradiation, 2D materials, Q-switched lasing.

**ABSTRACT:**

Van der Waals (vdW) heterostructures are receiving great attentions due to their intriguing properties and potentials in many research fields. The flow of charge carriers in vdW heterostructures can be efficiently rectified by the inter-layer coupling between neighboring layers, offering a rich collection of functionalities and a mechanism for designing atomically thin devices. Nevertheless, non-uniform contact in larger-area heterostructures reduces the device efficiency. In this work, ion irradiation had been verified as an efficient technique to enhance the contact and interlayer coupling in the newly developed graphene/$WSe_2$ heterostructure with a large area of 10 mm × 10 mm. During the ion irradiation process, the morphology of monolayer graphene had been modified, promoting the contact with $WSe_2$. Experimental evidences of the tunable interlayer electron transfer are displayed by investigation of photoluminescence and ultrafast absorption of the irradiated heterostructure. Besides, we have found that in graphene/$WSe_2$ heterostructure, graphene serves as a fast channel for the photo-excited carriers to relax in $WSe_2$, and the nonlinear absorption of $WSe_2$ could be effectively tuned by the carrier transfer process in graphene, enabling specific optical absorption of the heterostructure in comparison with separated graphene or $WSe_2$. On the basis of these new findings, by applying the ion beam modified graphene/$WSe_2$ heterostructure as a saturable absorber, Q-switched pulsed lasing with optimized performance has been realized in a Nd:YAG waveguide cavity. This work paves the way towards developing novel devices based on large-area heterostructures by using ion beam irradiation.

In parallel with research wave of graphene and other two-dimensional (2D) atomic crystals, heterostructures have attracted an increasingly attention over the past decade[1-3]. Heterostructures are layered materials bonded by the van der Waals (vdW) force, which are assembled by stacking different 2D materials on the top of each other[4]. The stacked layers with an artificial sequence can be empolyed to engineer the electronic band structures, allowing the interlayer electronic tunneling. The transition of electrons within an artificially designed energy barrier offers 2D heterostructures with exotic optical functionalities[5], including enhanced saturable absorption[6], ultrafast photo-responses[7], and modulated photoluminescence (PL)[8]. The performance of these unique incremental functions depends on the efficiency of electronic tunneling (i.e., the interlayer coupling) between the distinct 2D materials of the heterostructures[9-11].

However, as the layered materials are combined by the vdW force (considerably weak in comparison to the covalent bond), it is quite difficult to obtain an intense contact. Extra spaces, e.g. the trapped residual molecules and ripples (deformation) on 2D materials formed during the mechanical transfer or fabrication processes[12], also exist in the interlayer space. The loose contact reduces the interlayer coupling of heterostructures which hampers the development of large-area heterostructures. Recently, thermal annealing was used to drive the trapped residual molecules out of the interlayer space and thus improve the layers the interlayer coupling between the $MoS_2$ and $WS_2$ monolayers[13]. However, ripples (deformation) on 2D materials could not be flatten by simply annealing. It is urgent to seek efficient methods to minimize ripples for the further development and application of the heterostructure.

As one of the most intriguing applications in photonics, 2D materials have been widely applied in ultrafast laser generation as broadband saturable absorbers[14,15]. Due to different electronic structures of 2D materials, they have different performances of the saturable absorption. For instance, graphene has an ultrafast recovery time (relaxation time less than 150 fs) allowing a fast response in a laser system, but the optical modulation depth is low (around 1% for the monolayer graphene). Topological insulators (TI, e.g., $Bi_2Se_3$, $Bi_2Te_3$ and $Sb_2Te_3$) and transition metal dichalcogenides (TMDC, e.g. $WSe_2$, $WS_2$, $MoS_2$) have a high modulation depth, but the relaxation time is quite long (more than 500 fs) demonstrating a slow saturable absorber. One potential solution to complement advantages of TI (or TMDC) and graphene is to constitute heterostructure with graphene and other 2D materials. However, graphene has intrinsic ripples or corrugations because of thermal fluctuations[16], leading to the loose contact between neighbor layers and further hamper the interlayer coupling in the heterostructure. Recently, direct growing the thin nanocrystals on graphene substrate is reported as a method to constitute the heterostructure with a good interlayer coupling. But, this method only applies for limited materials and has stringent requirements for experimental operations. Besides, it cannot control the number of layers in the nanocrystal and the roughness of the heterostructure. Therefore, it is still desirable to explore new techniques to enhance interlayer coupling between graphene and 2D materials[6,17,18] for the further development of high-quality heterostructures.

The ion beam technology has been proved to be an effective way to modify the properties and functions of 2D materials. For example, ion irradiation could induce defects in 2D materials to realize effective tuning of their electronic structures and optical properties through carefully choosing irradiation conditions (e.g., ion species, energy, and fluence)[19-21]. In this work, we demonstrate the ion irradiation as an efficient tool to enhance the interlayer coupling, which enables efficient tailoring of optical functions of the heterostructures. The new-developed heterostructure used in this work consist of a monolayer graphene and $WSe_2$ (Tungsten-based dichalcogenide). By changing the fluence of ion beam, the average interlayer distance of the heterostructure is significantly decreased from ~4nm to ~0.2 nm with a minimal damage to structures of graphene and $WSe_2$. As the interlayer distance decreases, the interlayer coupling of the heterostructure has been enhanced considerably. Based on these findings, graphene/$WSe_2$ heterostructure is used as the saturable absorber for the Q-switched pulsed laser generation. Our work indicates ion irradiation as a novel way to tailor optical properties of the heterostructures.

The atomic force microscope (AFM) topographic data of the as-prepared graphene/WSe$_2$ heterostructure is shown in Figure 1a. As graphene is mechanically stacked onto monolayer WSe$_2$, there are ripples and corrugations on the transferred monolayer graphene due to the intrinsic thermal fluctuations and the transfer process (Figure 1b), resulting in a loose contact between layers (Figure 1c). Taking the thickness of the as-prepared heterostructure (~5 nm), graphene (~ 0.24 nm) and WSe$_2$ (~ 0.7 nm)[22,23] into account, the average interlayer distance is ~ 4 nm. After the carbon ion irradiation, the morphology of graphene/WSe$_2$ heterostructure could be modified (Figure 1d), leading to intense contact with the average thickness of 1.1nm. Considering the thickness of graphene and WSe$_2$ layers, the interlayer distance is determined to be ~0.2 nm after the ion irradiation, demonstrating the graphene/WSe$_2$ heterostructure has been compressed during the ion irradiation process (Figure 1e and 1f). More interestingly, the interlayer distance can be tailored by the fluence of the incident carbon ions. The interlayer distance has been gradually changed from 4 nm to 0.2 nm along with the increasing of the ion fluence (Figure 1g).

During the ion irradiation process, incident energetic ions pass through the heterostructure and have been injected into the substrate. Through the interaction with the heterostructure, the incident ions are slow down and transfer energy to the heterostructure. Meanwhile, a large number of near-surface atoms in the heterostructure are excited and obtain the momentum along with the direction of the incident ions. As a result, adjacent layers approach to each other, which decrease the interlayer distance of the heterostructure. In addition, the ripples and corrugations are transformed to tiny wrinkles after the irradiation. The evidence for the ion irradiation induced compression can be found in the Raman spectrum. The G-band of graphene represents the phonon modes between carbon atoms, which is sensitive to the state of stress, strain and the lattice disorder induced by compression. Figure 1h display the comparison of Raman signals of as-prepared and irradiated graphene/WSe$_2$ heterostructure. Raman peaks of graphene have the red shift after irradiation, demonstrating the graphene monolayer is compressed during the ion irradiation process.[24]

The quality of the heterostructure before and after the irradiation is also characterized by the Raman spectrum within the band of 100 nm–400 nm (WSe$_2$) and 1000 nm–3000 nm (graphene) in Figure 1h. For the WSe$_2$ layer, the position of the Raman peak (*P*) remains at 252.6 nm, while the bandwidth (*W*) of Raman signal of the as-prepared and irradiated ones is slight increased from 6.5cm$^{-1}$ (*S$_0$*) to 7.7 cm$^{-1}$ (*S$_3$*), indicatting after the irradiation, the intrinsic structure of WSe$_2$ were well-preserved. For graphene, the *G* and *2D* peaks in the as-prepared sample have the intensity ratio of 0.4 ($I_G/I_{2D}$), implying the nature of monolayer graphene[20]. The ratio of *D* and *G* peaks demonstrates the density of the zero-dimensional point-like defects in graphene. And the tiny value of *D/G* ratio also indicates the excellent quality of graphene in the as-prepared one[23]. Based on these results, one can conclude that both graphene and WSe$_2$ monolayer are not damaged during the irradiation.

According to previous discussion, the ion irradiation process can change the morphology of graphene, resulting in a close contact with the WSe$_2$ layer and consequently enhancement of interlayer coupling. This allows us to investigate the tuning effect of interlayer coupling, according to the measured optical functions of the heterostructures.

Figure 2a shows the normalized PL spectra emerged from the monolayer WSe$_2$ (orange line), heterostructures before (red line) and after ion irradiation (blue line), respectively. Dominated by the A excitation, the PL spectrum of the WSe$_2$ monolayer has a remarkable exciton emission at 748 nm, which is consistent with the previous study[25]. In contrast to the WSe$_2$, the PL spectrum of as-prepared heterostructure has little change in the shape, indicating an undisturbed energy state of the WSe$_2$ and a weak interaction between graphene and the monolayer WSe$_2$. After ion irradiation, the PL spectrum shows a remarkable band broadening and the blue shift (1 nm) of peak position. These notable variations may imply that the irradiation enhances the interlayer coupling and modifies the electronic structure of the graphene/WSe$_2$ heterostructure. Further evidence for the enhanced interlayer coupling is observed in the time-resolved PL decay kinetics characterization (Figure 2b) at 748 nm.

Before the irradiation, the as-prepared heterostructure has the same decay time as WSe$_2$ layer (2.45 ns), whilst the irradiated heterostructure possesses a faster decay (2.2 ns), suggesting an acceleration of charge transfer in WSe$_2$ due to the enhancement of interlayer coupling. Figure 2d and 2e are the PL mappings of the heterostructure before (Figure 2d) and after (Figure 2e) the ion irradiation at 748 nm.

In order to reveal the relationship between the interlayer coupling and the charge transfer in 2D materials, we have detected charge/excitons transfer of the irradiated heterostructure and two components (graphene monolayer and WSe$_2$ monolayer) through the pump - probe measurement by using a mode-locked Yb:KGW fiber laser (190 fs, 515 nm, 20 Hz)[26]. As shown in Figure 2f, the single-exponential decay with a time constant of 2.4ps/10ps indicates the ultrafast/slow carrier lifetime in graphene/WSe$_2$. More interestingly, similar experiment with the irradiated heterostructure (Figure 2f) exhibits two astonishing features. First, the magnitude of the signal (1.3) is 10 times higher than the monolayer WSe$_2$ and graphene, respectively. Second, the delay time (2.4 ps) is much shorter than the WSe$_2$ monolayer but close to that of graphene. These observations imply that the irradiated heterostructure possesses advantages of both graphene and WSe$_2$ monolayer due to the enhanced interlayer coupling. In respect to the isolated graphene/WSe$_2$, the heterostructure exhibits stronger and faster charge transfer effect.

To provide an insight into the observed phenomena, we performed first-principles calculations within the density-functional theory (DFT) of the graphene/WSe$_2$ heterostructure using the Vienna ab initio simulation package known as the VASP code[27-29]. The heterostructure is modeled by placing a graphene monolayer onWSe$_2$, as shown in Figure 3a. The lattice constant of the graphene/WSe$_2$ heterostructure is about 9.91 Å. The optimized equilibrium distance between graphene and WSe$_2$ is about 0.3 nm, close to the value of experiment. Figure 3b illustrates the electronic band structure and density of states (DOS) of the heterostructure. As one can see, both the gapless feature of graphene and semiconducting nature of WSe$_2$ are preserved in the heterostructure. The band hybridization between the graphene and WSe$_2$ in the valence and conduction band regions. The gapless graphene can provides a fast channel forrelaxation of photo-excited electrons in WSe$_2$, as illuminated by the energy diagram of the graphene/WSe$_2$ heterostructure (Figure 3c).

It is noteworthy that the graphene in our samples exhibits p-doping features due to the doping effect of the substrates, defects, water molecules, and oxygen in air. Meanwhile, WSe$_2$ has an intermediate states due to the *S*e-vacancies and a higher work function than graphene[30-32]. Under light illumination, electrons are excited into the intermediate states (1), with photo-generated holes left in the valence bands, creating excitons in WSe$_2$ monolayer. Through the interlayer coupling, the photo-generated electrons transfer from the intermediate states of WSe$_2$ to the conduction bands of graphene (2). The photo-excited excitons relax in both graphene and WSe$_2$ (3). Electrons transfer from the valence band of graphene to WSe$_2$ monolayer to keep the neutrality of WSe$_2$ (4). Consenquently, the graphene layer provides an additional fast channel for the recombination of photo-generated carriers in the WSe$_2$. Based on thisunique energy band structure of graphene/WSe$_2$ heterostructure, it is expected that the relaxation becomes faster in graphene/WSe$_2$ heterostructure,which is good agreement with experimental results in Figure 2b.

Faster relaxation in the heterostructure indicates potential applications in optoelectronic devices requiring ultrafast responses. Recently, the isolated WSe$_2$ and graphene have been reported to have the nonlinear saturable absorption[33-35], which can be used as modulators for the ultrafast laser emission. It is conceivable that the graphene/WSe$_2$ heterostructure has unique features of the saturable absorption compared with isolated components in virtue of the interlayer coupling. We have measured the nonlinear absorptions of 2D materials (WSe$_2$, graphene and graphene/WSe$_2$ heterostructure) by the Z-scan technology. As shown in Figure 4, the saturable absorption has been observed in all samples. In order to quantitatively determine the parameters of the saturable absorption, the nonlinear transmission is fitted by the equation below[31].

$$T(I) = 1 - \Delta T \times e^{-\frac{I}{I_{sat}}} - T_N \tag{1}$$

Where $T$ is the transmission, $I$ is the excitation energy, $T_N$ is the non-saturable absorbance, $\Delta T$ is the modulation depth, and $I_{sat}$ is the saturation intensity. In details, for monolayer graphene $I_{sat} \approx 0.86$ GW/cm$^2$ and $\Delta T \approx 0.36$ % (Figure 4a), whilst, for the monolayer WSe$_2$, $I_{sat} \approx 14.1$ GW/cm$^2$ and $\Delta T \approx 0.83$% (Figure 4b). The integration of distinct monolayers (graphene and WSe$_2$), ignoring the electronic tunneling, can be characterized by the product of their nonlinear transmission ($T_{graphene} \times T_{WSe2}$) without considering the interlayer coupling effect. As shown in Figure 4c, the saturation intensity (modulation depth) of this discrete integration is 9 GW/cm$^2$ (1.2 %). This processing of the nonlinear transmission is utilized as a reference to analyze the effect of electronic tunneling to the saturation intensity of the heterostructure.

The measured nonlinear transmission of the un-irradiated graphene/WSe$_2$ heterostructure is also shown in Figure 4c. In comparison to that from the integration of two monolayers (graphene and WSe$_2$), the measured optical transmission is faster to achieve the saturation indicating the weak interlayer coupling between layers. Considering the average interlayer distance of the heterostructure ~4 nm in Figure 1a, weak tunneling is attributed to the long interlayer distance. Figure 4d shows the nonlinear transmission of the irradiated heterostructure (6 MeV C$^+$ irradiation at the fluence of $1 \times 10^{13}$ ions/cm$^2$). After ion irradiation, the modulation depth and the saturation intensity are measured to be 2 % and 2.6 GW/cm$^2$, respectively. Compared with the integration of monolayers in Figure 4c, there is a great increasing/decreasing of the modulation depth/saturation intensity, suggesting strong interlayer coupling in the irradiated heterostructure. Figure 4e and f summarize the variation of the saturable absorption of the heterostructure as a function of the average interlayer distance. With the average thickness approaching 0.2 nm, the modulation depth and the saturation intensity have drastic variations. Compared with the monolayer WSe$_2$, the saturation intensity/modulation depth of the heterostructure is decreased/increased for 3.5/1.6 times.

Since the saturable absorption of the WSe$_2$ is enhanced by the integration with graphene monolayer, it is interesting to apply this new heterostructure for the generation of laser pulses. We therefore apply the irradiated graphene/WSe$_2$ heterostructures as the saturable absorber in yttrium aluminum garnet ceramic (Nd:YAG ceramic) waveguide[36], which is pumped by an 810-nm laser from a continuous-wave tunable Ti:Sapphire laser. As the pump power is above 100 mW, stable laser pulsesare generated by passive Q-switching. Figure 5b shows the pulse train of the output laser at the pump power of 600 mW. The pulse duration has a slow decreasing with the pumping power less than 400 mW. Above the pump power of 400 mW, the pulse duration is almost constant at 43.4 ns (Figure5c). Figure 5d depicts the peak power of the output laser as a function of the pump power. As one can see, the peak power has a maximum value of 0.69 W. Short pulse and higher peak power are the purpose of the development of the saturable absorber. In order to explore the advantages of the heterostructure for the Q-switching, we perform a similar experiment with the monolayer WSe$_2$ for comparison.

With a monolayer WSe$_2$ (instead of the heterostructure) as a saturable absorber, the pulse train is obtained under the same pumping condition (see the data of output power, the repetition rate and the pulse duration in Figure 5e-g). As one can see, for the WSe$_2$ based system, the maximum peak power is 0.38W, and the minimum pulse duration is 117 ns. The peak power/pulse duration is much higher/shorter than the one obtained from the heterostructure-based system. This clearly indicates the advantage of ion irradiated graphene/WSe$_2$ heterostructure over WSe$_2$ monolayer in photonic applications.

In addition to the WSe$_2$ monolayer, we have also designed novel structures to optimize optical properties of WSe$_2$ with multilayers. As shown in Figure 6a, graphene and WSe$_2$ monolayer overlap with each other, constituting multilayer heterostructures (G/W×2, G/W×3). In this structure, each graphene provides an additional channel for the neighboring WSe$_2$ layers to accelerate the optical response. Nonlinear absorptions of G/W×2, G/W×3 are displayed in Figure 6b and c. G/W×2 has the modulation depth and the saturation intensity of 2.8% and 2.2 GW/cm$^2$, and G/W×3 has values of 3.7 % and 4 GW/cm$^2$, respectively. Compared with the G/W×1 in Figure 4, both of the saturable absorption and the modulation depth have a nearly linear increasing in the multilayer heterostructure (Figure 6c), demonstrating the stack effect of graphene/WSe$_2$ heterostructure.

Carbon ion irradiation has significantly enhanced the interlayer coupling of new-developed large-area graphene/WSe$_2$ heterostructure (consisting of monolayer graphene and WSe$_2$), resulting in drastic variations of nonlinear optical properties of the heterostructures. When graphene/WSe$_2$ spacing is modified, heterostructure possesses larger modulation depth and lower saturation intensity than each separate 2D material due to the enhanced interlayer coupling. Consequently, new Q-switched lasers have been realized by using the modified graphene/WSe$_2$ heterostructure as saturable absorber, showing enhanced lasing performances in respect to the monolayer WSe$_2$-based system. This work provides a new strategy to evaluate the quality and tailor the optical parameters of 2D heterostructures (such as black phosphorus) for particular requirements of novel photonic and optoelectronics devices.

METHODS.

*Sample preparation.* The graphene/WSe$_2$ heterostructure used in this work is produced by the chemical vapor deposition (CVD) followed by wet-chemical transfer techniques. At first, the monolayer graphene and the monolayer WSe$_2$ are fabricated by CVD on copper and aluminum oxide substrates, respectively. Each monolayer has a dimension of 10 mm × 10 mm. As the WSe$_2$ monolayer with its pristine structure has low absorption in the near-infrared (NIR) region, we artificially tailor the atomic ratio of *W* and *Se* to 1.9 during the fabrication process, so that the intermediate states in the band gap of the WSe$_2$ can be generated to induce the NIR absorption. Afterwards, graphene is transferred onto the monolayer WSe$_2$ by the wet-chemistry transferchemical process. Due to vdW interactions, graphene and WSe$_2$ are bonded together to construct a graphene/WSe$_2$ heterostructure. Three CVD-prepared samples of graphene/WSe$_2$ heterostructures are irradiated by 6 MeV carbon (C$^{3+}$) ions at a fluence of $1\times10^{12}$ ions/cm$^2$, $5\times10^{12}$ ions/cm$^2$, and $1\times10^{13}$ ions/cm$^2$, respectively through a 2×1.7 MV tandem accelerator at Peking University.

***PL measurement.*** The steady-state PL and the time-resolved PL decay kinetics characterization are carried out via a PG2000-Pro-EX spectrometer. The excitation wavelength is 400 nm from the second harmonic output of a Ti:sapphire laser (Maitai HP, Spectra-Physics) at 80 MHz. The time-resolved PL measurement is performed by the time-correlated single photon-counting (TCSPC) technique and the data are collected by a Halcyone spectrometer (Ultrafast System).

***Pump-probe measurement.*** During the measurement, the laser pulse is divided into a probe beam and a pumping beam via a beam splitter. The peak power of the probe beam is modulated as approximately 8% of the pump beam through a system of half plate and polarizer, in order to remove the nonlinear effect caused by the probe beam. The polarization of pump and probe beam is set at the magic angle to avoid anisotropic effect. The waist of the probe beam is 22 mm, which was smaller than the pump waist (152 mm). Meanwhile, we put a variable delay into the pump path and recorded the variation of the probe beam intensity versus the delay time by an energy detector after the pump beam.

***DFT calculations.*** Our first-principles calculations were performed within density-functional theory (DFT) using the Vienna ab initio simulation package known as the VASP code[27-29]. The electronic-ion interaction was described by projector augmented wave method (PAW)[37,38]. The energy cutoff of the plane waves was set to 450 eV with an energy precision of 10$^{-5}$ eV. The electron exchange–correlation function was treated using a generalized gradient approximation (GGA) in the form proposed by Perdew, Burke, and Ernzerhof (PBE)[39].The vdW interactions is described by DFT-D2 method of Grimme[40].The Monkhorst-Pack $k$-point meshes[41] for the Brillouin zone (BZ) sampling used in structural optimization and electronic structure calculations are 3×3×1 and 5×5×1, respectively. A vacuum region up to 15 Å was applied along the z-direction to exclude the interaction between adjacent images. Both atomic positions were fully optimized using the conjugate gradient (CG) algorithm until the maximum atomic forces were less than 0.1eV/Å.The full relaxed lattice constant for hexagonal $WSe_2$ and graphene is 3.316Å and 2.469Å, which is consistent with previous value 3.32Å and 2.44Å, respectively.[42,43] The graphene/$WSe_2$ heterostructure is modeled by placing a 4×4 graphene on a 3×3 $WSe_2$ monolayer. The slight lattice mismatch ~0.67% between these two lattices is negelected.

ASSOCIATED CONTENT

AUTHOR INFORMATION

**Corresponding Author**

*To whom correspondence should be addressed. E-mail: drfchen@sdu.edu.cn; zmw@sdu.edu.cn; hzhang@szu.edu.cn.**Author Contributions**

The manuscript was written through contributions of allauthors. All authors have given approval to the final version ofthe manuscript.

**Notes**

The authors declare no competing financial interest.

ACKNOWLEDGMENT

This work is supported by the National Natural Science Foundation of China (Nos. 11535008, 21433006, and 61435010). Y. T. acknowledges the financial support from Young Scholars13

Program of Shandong University (No. 2015WLJH20). Y. T. thanks suggestions from Prof. Yinglin Song.

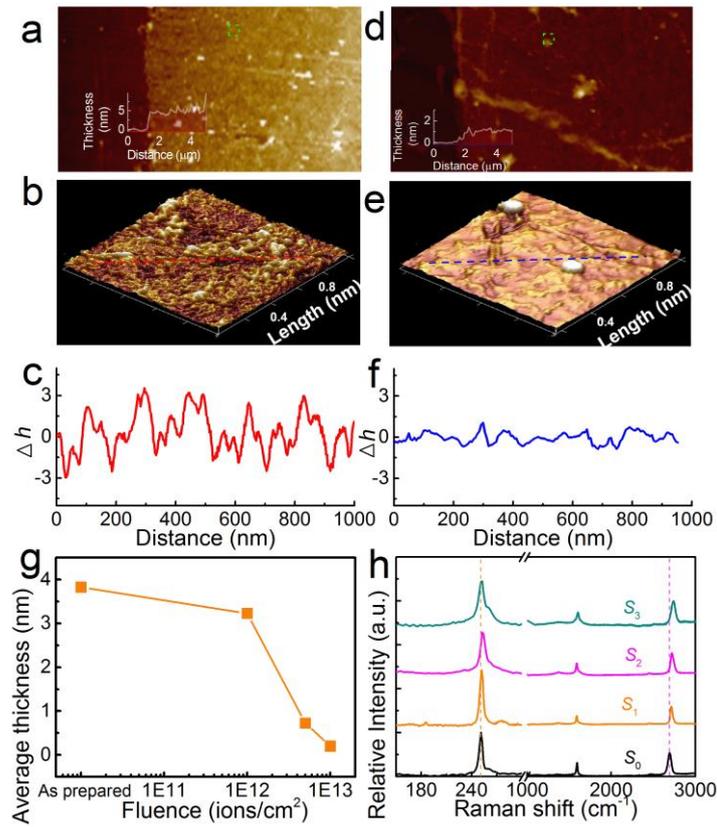

**Figure 1.** Morphology of the graphene/WSe$_2$ heterostructure before a) and after d) the ion irradiation captured by the AFM. b) and e) are the enlarged views of green squares in a) and d). c) and f) are the thickness variation along the red and blue lines in b) and e). g) The variation of the thickness of the graphene/WSe$_2$ heterostructures as a function of the fluence of the carbon ion beam. h) is Raman spectrum of the graphene/WSe$_2$ heterostructure before and after the ion irradiation.

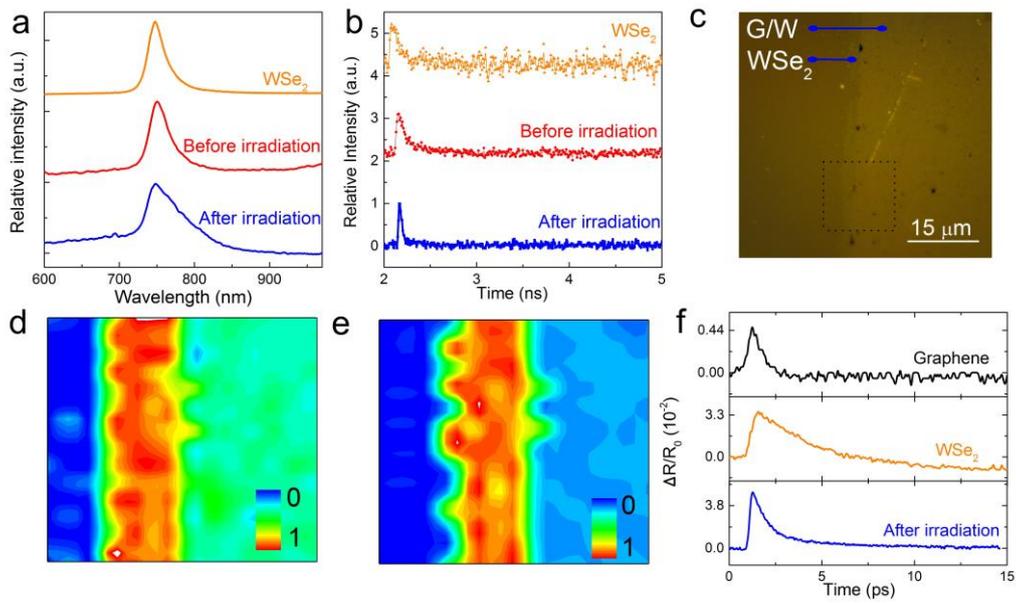

**Figure 2.** a) PL spectra, b) PL decay curves, c) Microscope photo of the G/W heterostructure. PL mappings of G/W heterostrucure in the black square of c) before d) and after e) ion irradidation. f) single-exponential decay curves of WSe$_2$, graphene/WSe$_2$ heterostructure before and after the ion irradiation.

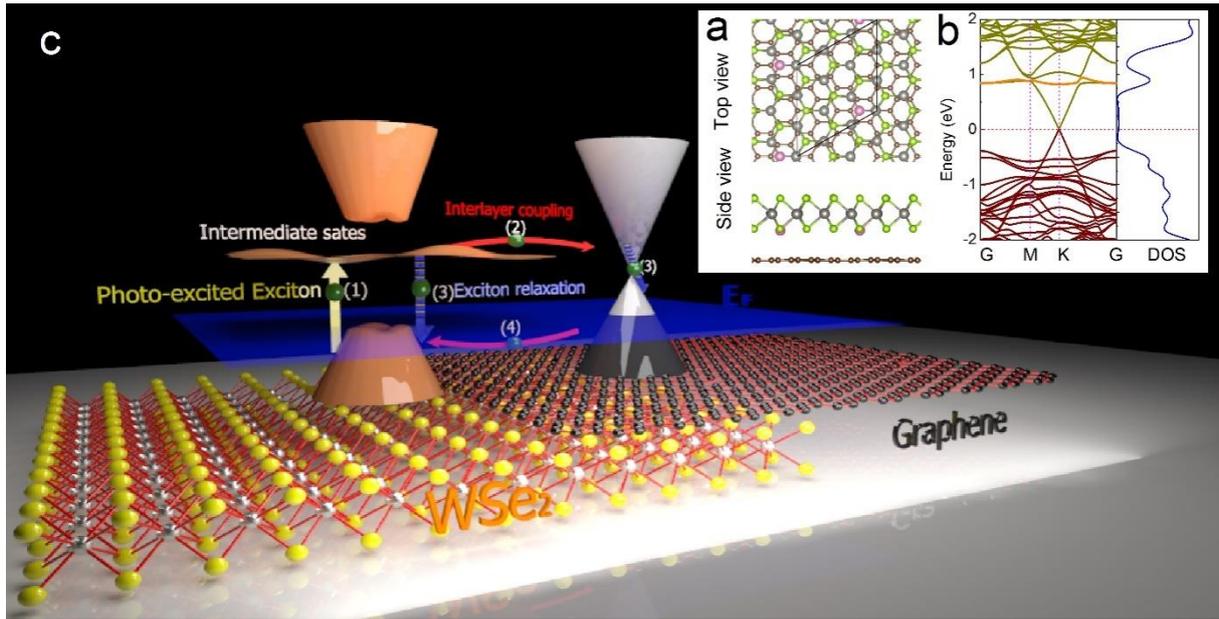

**Figure 3.** a) Geometry (top and side view) and b) electronic band structure and DOS of the graphene/WSe$_2$ heterostructure. The flat band marked in orange indicates the intermediate states induced by *Se*-vacancies. c) Schematic diagram showing electronic transitions in graphene/WSe$_2$ heterostructure.

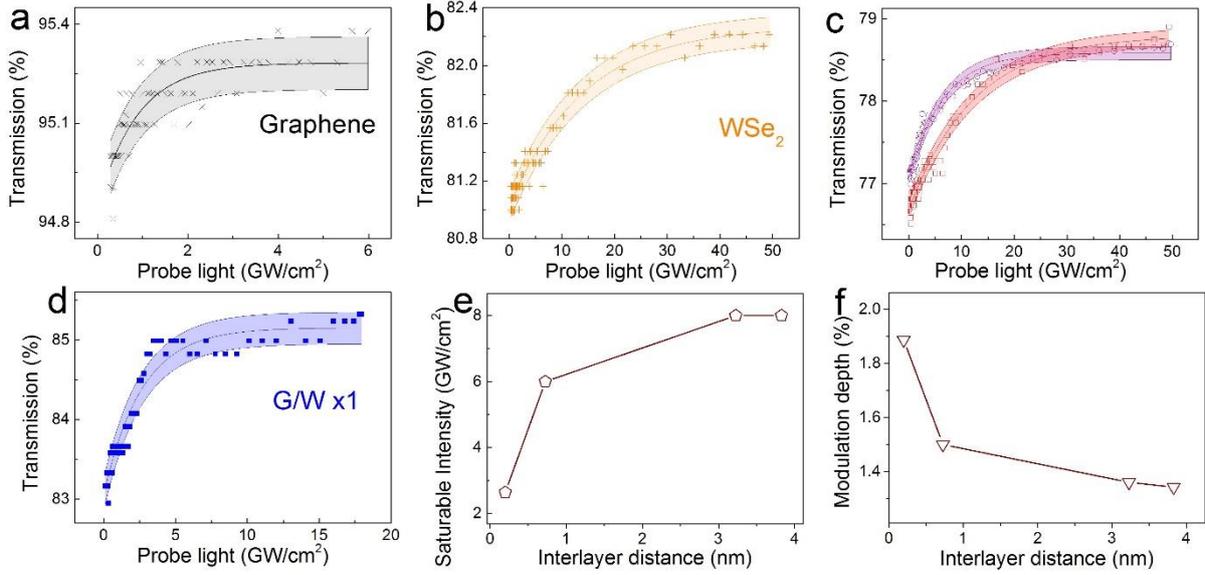

**Figure 4.** Nonlinear transmission of a) monolayer graphene, b) monolayer WSe$_2$, c) as-prepared graphene/WSe$_2$ heterostructure, and d) irradiated graphene/WSe$_2$ heterostructure measured by the Z-scan technology at the wavelength of 1064 nm. The nonlinear transmission product of monolayer graphene and monolayer WSe$_2$ are shown in c) as Cyan squares. Variation of the modulation depth e) and the saturation intensity f) of the graphene/WSe$_2$ heterostructure along with the thickness of the graphene/WSe$_2$ heterostructure.

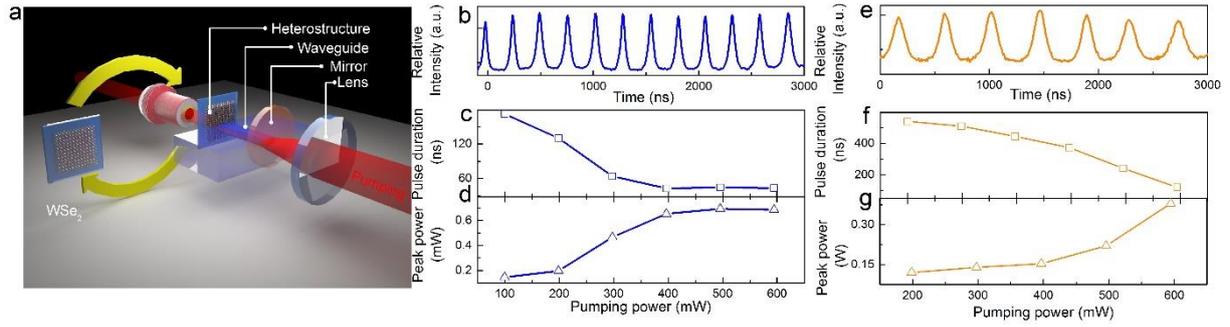

**Figure 5.** a) Experimental setup for the Q-switched waveguide laser. The b) pulse train, c) output power, and d) repetition rate and pulse duration of the output laser modulated by the graphene/WSe$_2$ heterostructure. The e) pulse train, f) output power, g) repetition rate and pulse duration of the output laser modulated by the monolayer WSe$_2$.